\documentclass[12pt]{article}
\usepackage{hyperref}
\usepackage{graphicx}
\usepackage{color}
\usepackage{amsmath,amssymb} % Symbols for itemise
\definecolor{myblue}{rgb}{0.14,0.11,0.49}
\definecolor{myred}{rgb}{0.74,0.22,0.15}
\definecolor{mygreen}{rgb}{0.05,0.52,0.42}
\definecolor{myyellow}{rgb}{0.96,0.92,0.13}
\definecolor{myorange}{rgb}{1,0.61,0.36}
\definecolor{mypurple}{rgb}{0.71,0.02,1}
\definecolor{noir}{gray}{0.} % black

\newcommand{\Couleur}[1]{\textcolor{noir}{#1}}

\definecolor{htc}{rgb}{1,1,1} % heading text colour
\newcommand{\Mat}[1]{{{\boldsymbol{#1}}}}

\def\be{\begin{equation}}
\def\ee{\end{equation}}
\def\bea{\begin{equationarray}}
\def\eea{\end{equationarray}}
\def\bc{\begin{center}}
\def\ec{\end{center}}
\def\bi{\begin{itemize}}
\def\ei{\end{itemize}}
\def\bs{\begin{slide}}
\def\es{\end{slide}}
\def\dd{\mathrm{d}}

\def\noi{\noindent}
\title{On continuum dynamics and the electromagnetic field in the scalar ether theory of gravitation}
\author{
Mayeul Arminjon\\
\small\it Univ. Grenoble Alpes, CNRS, Grenoble INP 
, 3SR, F-38000 Grenoble, France
} 

\date{}
\begin{document}
%%%%%%%%%%%%%%%%%%%%%%%%%%%%%%%%%%%%%%%%%%%%%%%%%%%%%%%

\maketitle

\begin{abstract}
\noindent We summarize a recent work on the title subject, skipping the detailed calculations but introducing the basic points with enough detail. The theory considered is formulated in a preferred reference frame in a four-dimensional spacetime endowed with a curved ``physical" metric. The dynamics of a test particle is defined by an extension of the special-relativistic form of Newton's second law. This determines the dynamical equation verified by the energy-momentum tensor ${\bf T}$ of a ``dust" continuum, also in the presence of a non-gravitational external force. The dynamical equation for ${\bf T}$ thus obtained is assumed valid for a general continuous medium or a system of fields. When the non-gravitational force is the Lorentz force, this equation in turn determines the Maxwell equations in a  gravitational field for the present theory. They are consistent with the dynamics of photons i.e. with the geometrical optics of the theory. Except for a constant gravitational field, they seem to imply some local production or destruction of electric charge. The possible amounts are yet to be assessed.
%\vspace{30mm}

\end{abstract}
%%%%%%%%%%%%%%%%%%%%%%%%%%%%%%%%%%%%%%%%%%%%%%%%%%%%%%%%%%%%%%%%%%%%%%%%%%%%%%%%
\section{Introduction}
%%%%%%%%%%%%%%%%%%%%%%%%%%%%%%%%%%%%%%%%%%%%%%%%%%%%%%%%%%%%%%%%%%%%%%%%%%%%%%%%

A theory based on an alternative view of gravity has been proposed and investigated. The idea of the theory lies in a semi-heuristic interpretation of gravitation as a pressure force in an imagined perfect fluid or ``ether" and in the introduction of a finite velocity of propagation through a compressibility of the ether \cite{A8}. 
\footnote{\label{other works}
Our approach differs in spirit from numerous works (some of them rather recent), in which it is aimed at deriving classical equations of electromagnetism and/or gravitation from some mechanical model of ether. For instance, Zareski \cite{Zareski2001} considered an ether being a special type of elastic solid and found that the equations governing this medium are identical in form with the classical Maxwell equations. Later on, he investigated a generalization of this model in which Einstein gravity is assumed to be present %; according to him, at a short distance the gravitational attraction would dominate over the e.m. force in this model, and could represent the strong interaction 
\cite{Zareski2014}. An elastic-solid ether and its relation with electromagnetism and gravitation were also studied in some detail by Dmitriyev \cite{Dmitriyev1992, Dmitriyev2004}. In our approach, the interpretation of gravity as a pressure force in a compressible fluid leads us to state a self-consistent set of equations which builds a new theory of gravitation. %In contrast with the former works \cite{Zareski2001}--\cite{Dmitriyev2002}, 
The Maxwell equations discussed in the present paper are derived, not from some assumed mechanical constitutive relation in the ether, but from continuum dynamics applied to a material medium having a charge density and being subjected to the Lorentz force.
}
The construction of a theory based on this idea and accounting for special relativity, leads to a scalar theory written in a preferred reference frame in a four-dimensional spacetime endowed with a flat ``background" metric and a curved ``physical" metric \cite{A9,O3,A35}. Thus, although in its concept this is a very different theory from general relativity (GR) and from most alternative theories, nevertheless this theory also considers a curved Lorentzian spacetime as the main mathematical tool. (The curvature scalar and tensors are not needed, however.) Also like GR, this theory coincides with special relativity if the gravitational field vanishes. 

The aim of this paper is to summarize the development \cite{A54} of the theory to describe continuum dynamics and the electromagnetic (e.m.) field in the presence of gravitation. The equations of the e.m. field are indeed obtained as an application of the equation for continuum dynamics in the presence of gravitation and a non-gravitational external field.

In contrast with \cite{A54}, the present paper does not discuss the motivations for the theory nor its experimental check, and it does not present in detail the calculations related either to Newton's second law in a general continuum or to the e.m. field. However, it summarizes the results while exposing with enough detail the basic points: e.g., the general notion of a reference fluid, the associated physical space, and the local time along a trajectory when such a reference fluid is given; Newton's second law for a test particle in a general reference fluid in a curved spacetime; and the transition from Newton's second law for a test particle to the dynamical equation satisfied by a dust continuum.

%%%%%%%%%%%%%%%%%%%%%%%%%%%%%%%%%%%%%%%%%%%%%%%%%%%%%%%%%%%%%%%%%%%%%%%%%%%%%%%%
%%%%%%%%%%%%%%%%%%%%%%%%%%%%%%%%%%%%%%%%%%%%%%%%%%%%%%%%%%%%%%%%%%%%%%%%%%%%%%%%
\section{Assumptions used in this work}
%%%%%%%%%%%%%%%%%%%%%%%%%%%%%%%%%%%%%%%%%%%%%%%%%%%%%%%%%%%%%%%%%%%%%%%%%%%%%%%%

We note first that the equation for the scalar field (which is the flat-spacetime wave equation \cite{A35}) is {\it not} used in this work. In fact, the very existence of the flat background spacetime metric is not needed, except for the comparison with the geometrical optics i.e. the transition to the dynamics of a free photon (Subsect. \ref{photons}).

\subsection{Assumptions common with general relativity}

Two assumptions of this work are common with GR:
\bi
\item {\it a}) Our space and time measurements may be arranged so as to be described by a Lorentzian metric \Couleur{$\Mat{\gamma} $}, i.e., a metric with \Couleur{$(+ - - -)$} signature on a 4-dimensional \hypertarget{Medium-by-T}{manifold} \Couleur{$\mathrm{V}$} (the spacetime). \\

\item {\it b}) A continuous medium or a continuous field is defined by the expression of its energy-momentum tensor \Couleur{${\bf T}$}, which is a spacetime tensor depending on some state variables. The dynamical laws can be universally expressed in terms of \Couleur{${\bf T}$}, independently of the specific medium/field. To introduce this tensor, we follow Fock \cite{Fock1964}: there is no need for a Lagrangian. (This point is explained in detail in Ref. \cite{A54}: Assumption ({\it b}) in Sect. 2 there.)\\

\ei 
\noi Moreover, several notions used in this theory can also be defined in GR:

\bi

\item i) The notion of a reference fluid. This is a 3-D network \Couleur{$\mathcal{F}$} of {\it reference points,} each of them being defined by its world line. Each reference world line has to be time-like for an admissible reference fluid \cite{Cattaneo1958}. An admissible reference fluid \Couleur{$\mathcal{F}$} may thus be defined by the associated unit tangent 4-vector field on $\mathrm{V}$, \Couleur{$\Mat{U}=\Mat{U}_\mathcal{F}$}  (e.g. Ref. \cite{RodriguesCapelas2007}): the reference world lines are the integral curves of \Couleur{$\Mat{U}_\mathcal{F}$}. For a given reference fluid \Couleur{$\mathcal{F}$}, a system of {\it adapted coordinates} on $\mathrm{V}$, or an adapted chart of $\mathrm{V}$, is a system of spacetime coordinates $x^\mu \ (\mu=0,...,3)$ for which the vector of the spatial coordinates, \Couleur{${\bf x} \equiv (x^i) \in \mathbb{R}^3$}, is constant on any reference world line (e.g. Refs. \cite{Moller1952,Cattaneo1969}). The notion of a reference fluid was thus first defined for GR, but it is not commonly used explicitly there, although coordinate systems are of course routinely used, and the data of one coordinate system does define a reference fluid (in the domain of definition of the coordinate system) --- namely, the one whose reference world lines are defined to be the lines ${\bf x}\equiv (x^i)=\mathrm{Constant}$.\\

\item ii) Synchronized local time. Given an admissible reference fluid $\mathcal{F}$, consider an open curve in spacetime, defined as a one-to-one smooth mapping $C: \xi \mapsto C(\xi )=X\in\mathrm{V}$ from a real interval $\mathrm{I}$ into the spacetime. Up to a constant shift, we can define uniquely the {\it synchronized local time} on this trajectory. This is the proper time measured at the successive reference points of $\mathcal{F}$ met along the trajectory, synchronized along that trajectory according to the (Poincar\'e-Einstein) standard synchronization procedure in $\mathcal{F}$ \cite{Cattaneo1958, L&L, A16}. Because in an adapted chart a reference point of $\mathcal{F}$ can be defined by the vector \Couleur{${\bf x} \in \mathbb{R}^3$}, we denote the synchronized local time by $t_{\bf x}(\xi )$ --- even though in general the different values of $\xi $ correspond with different reference points ${\bf x}={\bf x}(\xi )$, unless the trajectory in question is a reference world line. Explicitly \cite{Cattaneo1958, A16}:
\be\label{dt_x} 
\frac{\dd t_{\bf x}}{\dd \xi} = \frac{\sqrt{\gamma _{00}}}{c} \left( \frac{\dd x^0}{\dd \xi} + \frac{\gamma _{0i}}{\gamma _{00}}\,\frac{\dd x^i}{\dd \xi} \right ).
\ee
({\it We are considering adapted coordinates henceforth.}) The sense in which $t_{\bf x}(\xi )$ is indeed the standardly synchronized proper time along the trajectory is explained by Landau \& Lifshitz \cite{L&L}, after Eq. (88.9). (The line of reasoning there does not depend on the assumption of stationarity.) Note that (\ref{dt_x}) is invariant under any regular coordinate change for which the spatial coordinate change does not depend on the time coordinate \cite{Cattaneo1958}, hence it is well defined for a given reference fluid $\mathcal{F}$ and a given curve $C$.\\

\item iii) Space manifold associated with a reference fluid. This notion too can be defined in GR also but, to our knowledge, it has not been used yet. The ``physical space" associated with a reference fluid \Couleur{$\mathcal{F}$} can be defined to be the set of the reference world lines of \Couleur{$\mathcal{F}$} \cite{A16}. It thus depends on the reference fluid which is considered. Any non-vanishing vector field \Couleur{$\Mat{U}$} on \Couleur{$\mathrm{V}$} (whether \Couleur{$\Mat{U}$} is time-like or not) defines a unique reference fluid: the one whose reference world lines are the maximal integral curves of \Couleur{$\Mat{U}$} \cite{A52}. The set of these curves is denoted by \Couleur{$\mathrm{N}_\Mat{U}$}. (These curves, hence that set, are not changed if $\Mat{U}$ is replaced by $\Mat{U}'\equiv \lambda \Mat{U}$ with $\lambda $ any smooth, non-vanishing real function defined on $\mathrm{V}$ \cite{CalcutGompf2013}.) If \Couleur{$\Mat{U}$} is ``normal" in a precise sense, then adapted charts do exist for that reference fluid \Couleur{$\mathcal{F}_\Mat{U}$} \cite{A52}. They allow one to endow the space \Couleur{$\mathrm{N}_\Mat{U}$} with a canonical structure of {\it 3-D differentiable manifold} \cite{A52}. So, given a reference fluid, we can define {\it spatial fields,} i.e., fields over the space manifold \Couleur{$\mathrm{N}_\Mat{U}$} --- either scalar, vector, or tensor fields. In particular, the spatial metric defined from \Couleur{$\Mat{\gamma} $} by Landau \& Lifshitz \cite{L&L} or M\o ller \cite{Moller1952} is a \Couleur{$(0\ 2)$} spatial tensor field \Couleur{$\Mat{g} =\Mat{g}_\Mat{U}$} (it depends on the reference fluid). It makes \Couleur{$\mathrm{N}_\Mat{U}$} a Riemannian manifold --- but the spatial metric \Couleur{$\Mat{g}_\Mat{U}$} depends on time.\\

\item iv) Globally-synchronized reference fluid. We shall consider a special admissible reference fluid \Couleur{$\mathcal{E}$}, with four-velocity vector field \Couleur{$\Mat{U}_\mathcal{E}$}, and we shall assume that it is globally synchronized. By this, we mean that there is a {\it global} space-time coordinate system --- i.e., the mapping \Couleur{$X \mapsto (x^\mu )$} is defined for any event $X$ --- which is adapted to \Couleur{$\mathcal{E}$}, and in which the components of the spacetime metric \Couleur{$\Mat{\gamma}$} verify
\be\label{gamma_0i=0}
\Couleur{\gamma_{0 i}=0} \quad (i=1,2,3).
\ee
We note that the  existence of a  {\it local} chart (defined in a neighborhood of an arbitrary given event), in which (\ref{gamma_0i=0}) is true, is warranted in a generic spacetime (e.g. Ref. \cite{L&L}). Condition (\ref{gamma_0i=0}) alone does not specify a unique reference fluid, even less a unique coordinate system. The preferred character of the reference fluid $\mathcal{E}$ assumed in the theory appears with the dynamics.

%%%%%%%%%%%%%%%%%%%%%%%%%%%%%%%%%%%%%%%%%%%%%%%%%%%%%%%%%%%%%%%%%%%%%%%%%%%%%%%%
\ei
%%%%%%%%%%%%%%%%%%%%%%%%%%%%%%%%%%%%%%%%%%%%%%%%%%%%%%%%%%%%%%%%%%%%%%%%%%%%%%%%
\subsection{Dynamics of a test particle based on Newton's second law}
%%%%%%%%%%%%%%%%%%%%%%%%%%%%%%%%%%%%%%%%%%%%%%%%%%%%%%%%%%%%%%%%%%%%%%%%%%%%%%%%
\bi
\item i) Newton's second law in a curved spacetime. The original dynamics assumed in this theory is a central point of the latter. We assume an extension to a general Lorentzian spacetime \cite{A16} of the special-relativistic form of Newton's second law. It may be defined for any admissible reference fluid, and it too depends on the reference fluid. It has the following form: 
\be\label{Newton 2nd law}
\Couleur{{\bf F} + (E/c^{2}) {\bf g} =  D{\bf P}/Dt_{\mathbf{x}}}.
\ee
In this equation, \Couleur{${\bf F}$} is the non-gravitational force; \Couleur{${\bf g}$} is the gravity acceleration;
\Couleur{$E$} is the energy of the test particle: \Couleur{$E = m(v) c^{\, 2}$} for a mass point, with
\be\label{m(v)}
\Couleur{m(v) \equiv  m_0\gamma_{v} ,\qquad \gamma_{v}\equiv 1/\sqrt{1-(v^{ 
2}/c^{2})}}.
\ee
This is the rest-mass plus kinetic energy and it may be called the ``purely material" energy of the particle, for it does not take into account the potential energy of the particle in either the gravitational field or the electromagnetic field. Further, \Couleur{${\bf v}$} is the 3-velocity (relative to the admissible reference fluid \Couleur{$\mathcal{F}$} which is considered, the latter being characterized by the four-velocity field $\Mat{U}$): it is  measured with the synchronized local time \Couleur{$t_{\mathbf{x}}$}, and its modulus \Couleur{$v$} is defined with the space metric \Couleur{$\Mat{g}=\Mat{g}_\Mat{U}$}:
\be\label{Def v}
\Couleur{v^{\, i} \equiv \frac{\dd x^i}{\dd t_{\mathbf{x}}}, \quad v \equiv  [\Mat{g}({\bf v}, {\bf v})]^{1/2} =(g_{ij} v^i\, v^j)^{1/2}}.
\ee 
In the case of a synchronized reference fluid, thus if Condition (\ref{gamma_0i=0}) is satisfied, the synchronized local time (\ref{dt_x}) has a simple relation to the coordinate time $t\equiv x^0/c$ \cite{A16}:
\be
\dd t_{\bf x}/\dd t = \sqrt {\gamma_{00}} \equiv \beta(t,{\bf x}).
\ee
In that case, it suggests itself to parameterize the trajectory of the particle with the coordinate time $t$ and we have  
\be
v^i(t)=\frac{1}{\beta(t,{\bf x}(t))}\,\frac{\dd x^i}{\dd t}. 
\ee
Still, \Couleur{${\bf P }\equiv  (E/c^{\, 2}) {\bf v}$} is the momentum, and \Couleur{$D/Dt_{\mathbf{x}} $} is the relevant time-derivative of a vector on a curve in the space manifold \Couleur{$\mathrm{N}_\Mat{U}$} endowed with the time-dependent spatial metric \Couleur{$\Mat{g}=\Mat{g}_\Mat{U}$} \cite{A16}. It is given by
\be\label{Def D/Dt}
\frac{D{\bf P}}{Dt_{\mathbf{x}}} \equiv  \frac{\dd t}{\dd t_{\bf x}} \left(\frac{D_{0\, }{\bf P}}{Dt}+ \frac{1}{2}{\bf t.P}\right ), 
\quad {\bf t} \equiv  \Mat{g}^{-1}{\rm {\bf .}}\frac{\partial {\kern 1pt}{\kern 
1pt}{\kern 1pt}\Mat{g}}{\partial \,t},
\ee
where $D_{0\, }/\textit{Dt}$ is the absolute derivative relative to the space metric of the time $t$ at which the derivative is to be calculated; explicitly:
\be\label{D_0 P/Dt}
\left ( \frac{D_0{\bf P}}{Dt} \right )^i \equiv \frac{\dd P^i}{\dd t} + \Gamma ^i_{jk} P^j\, \frac{\dd x^k}{\dd t},
\ee
where the $\Gamma ^i_{jk}$'s are the Christoffel symbols of the metric $\Mat{g}$ of the time $t$. [The definition (\ref{Def D/Dt}) does not need that a synchronized reference fluid is considered.]\\

\item ii) The gravity acceleration. Newton's second law (\ref{Newton 2nd law}) is compatible with the geodesic motion of GR, provided a peculiar (velocity-dependent) form  of the gravity acceleration  \Couleur{${\bf g}$} is assumed \cite{A16}. However, we assume a different, simpler form \cite{A15}:
\be\label{Def vector g}
\Couleur{{\rm {\bf g}}=-\,c^{2}\,\frac{\mbox{grad}_{g} {\kern 1pt}\beta }{\beta },
\quad
\left( {\mbox{grad}_{g} \beta } \right)^{i}\equiv g^{ij}\beta_{,j} \quad
\left( {g^{ij}} \right)\equiv \left( {g_{ij}} \right)^{-1}}.
\ee
(Recall \Couleur{$\beta \equiv \sqrt {\gamma_{00} }$}.) This form came out from a mechanism/interpretation of gravity as a pressure force \cite{A8,A9}. It can also be {\it derived} by demanding that (i) the metric field \Couleur{$\Mat{\gamma} $} should be a spatial potential for a space vector \Couleur{{\bf g}}, and (ii) the law of motion (\ref{Newton 2nd law}) should imply geodesic motion for ``free" test particles in a \textit{static} metric \cite{A16}. The form (\ref{Def vector g}) is valid only in the preferred reference fluid $\mathcal{E}$. It is stable under the coordinate transformations (either global or local) that are internal to  $\mathcal{E}$ and preserve the synchronization condition (\ref{gamma_0i=0}). The general form of these coordinate changes is \cite{A16}:
\be\label{spatial change +f(t)}
x'^0 = \varphi (x^0), \quad x'^i= \psi ^i(x^1,x^2,x^3).
\ee

\vspace{1mm}
\item iii) Four-acceleration of a mass test particle. This is defined to be the absolute derivative of the four-velocity, relative to the spacetime metric (e.g. \cite{L&L}). Using Newton's second law (\ref{Newton 2nd law}) with the gravity acceleration (\ref{Def vector g}) and the explicit definition (\ref{Def D/Dt}) of the derivative \Couleur{$D {\bf P}/Dt_{\mathbf{x}}$}, we can compute \cite{A54} 
\be\label{4-accel}
\Couleur{A^0 = \frac{1}{2 \beta ^2}\,g_{jk,0}\,U^j\,U^k +  \frac{\gamma _v}{\beta }\frac{{\bf F} {\bf .v}}{m_0\,c^3}},\qquad
\Couleur{A^i = \frac{1}{2}\,g^{ij} g_{jk,0}\,U^0\,U^k +\gamma _v\frac{ F^i}{m_0\,c^2}}.
\ee
Here \Couleur{${\bf U}$}, with components \Couleur{$U^\mu $}, is the 4-velocity of the mass point. (This 4-velocity \Couleur{${\bf U}$} should not be confused with \Couleur{$\Mat{U}$}, that is the 4-velocity field of the preferred reference fluid.)\\

\ei

%%%%%%%%%%%%%%%%%%%%%%%%%%%%%%%%%%%%%%%%%%%%%%%%%%%%%%%%%%%%%%%%%%%%%%%%%%%%%%%%
\section{Continuum dynamics}
%%%%%%%%%%%%%%%%%%%%%%%%%%%%%%%%%%%%%%%%%%%%%%%%%%%%%%%%%%%%%%%%%%%%%%%%%%%%%%%%
%%%%%%%%%%%%%%%%%%%%%%%%%%%%%%%%%%%%%%%%%%%%%%%%%%%%%%%%%%%%%%%%%%%%%%%%%%%%%%%%
\subsection{Four-acceleration and dynamical equation for a dust continuum}
%%%%%%%%%%%%%%%%%%%%%%%%%%%%%%%%%%%%%%%%%%%%%%%%%%%%%%%%%%%%%%%%%%%%%%%%%%%%%%%%
``Dust" is defined to be a continuum made of coherently moving, non-interacting particles, each of which conserves its rest mass. Thus, Newton's second law (\ref{Newton 2nd law}) applies to each individual particle which constitutes that continuum. The energy-momentum tensor for the dust continuum is \cite{Fock1964}:
\be\label{T dust}
\Couleur{T^{\, \mu \nu } = \rho ^\ast \,c^{\, 2}\,U^{\, \mu }U^{\, \nu } },
\ee
where $\rho ^\ast $ is the rest-mass density field in the rest frame of the continuum. (The \Couleur{$U^\mu $}'s are now the components of the 4-velocity field of the dust continuum.) Mass conservation writes: \Couleur{$(\rho^\ast U^{\nu })_{\,;\nu}=0$} \cite{Fock1964}. With (\ref{T dust}), we get thus:
\be\label{div-T dust}
\Couleur{T^{\mu \nu}_{\ \,;\nu} = \rho^\ast c^2\, U^{\mu }_{\,;\nu}\,U^\nu = \rho ^\ast  c^2\,A^\mu}.
\ee
The four-acceleration (\ref{4-accel}) deduced from Newton's second law (\ref{Newton 2nd law}) is valid, as is the latter, for the constitutive particles of the dust. To write this for the dust when it is seen as a continuum, we substitute into (\ref{4-accel}) ${\bf F}={\bf f} \delta V$ and $m_0 = \rho _0 \delta V$, with ${\bf f}$ and $\rho _0$ the volume density fields of the non-gravitational external force and the rest-mass, respectively. ($\delta V\equiv \sqrt{g} \,\dd x^1 \dd x^2 \dd x^3$, with $g\equiv \mathrm{det} (g_{ij})$, is the volume element of the continuum, as measured from the reference fluid $\mathcal{E}$.) Inserting this modified version of (\ref{4-accel}) into (\ref{div-T dust}), and accounting for the relation $\rho _0 = \gamma _v \rho ^\ast $ \cite{L&L,A15}, we get:
\be\label{Eq T-f}
\Couleur{T^{0 \nu}_{\ \,;\nu} =b^0({\bf T})+\frac{{\bf f.v}}{c\beta}},
\qquad \Couleur{T^{i \nu}_{\ \,;\nu} =b^i({\bf T})+ f^i},
\ee
where
\be\label{b^mu}
\Couleur{b^0({\bf T}) \equiv \frac{1}{2\beta^2}\,g_{ij,0}\,T^{ij}},
\quad \Couleur{b^i({\bf T}) \equiv \frac{1}{2}\,g^{ij}g_{jk,0}\,T^{0k}}.
\ee

%%%%%%%%%%%%%%%%%%%%%%%%%%%%%%%%%%%%%%%%%%%%%%%%%%%%%%%%%%%%%%%%%%%%%%%%%%%%%%%%
\subsection{Induction from a dust to a general continuum}
%%%%%%%%%%%%%%%%%%%%%%%%%%%%%%%%%%%%%%%%%%%%%%%%%%%%%%%%%%%%%%%%%%%%%%%%%%%%%%%%

Dynamical equation for a general continuum: We assume that the same dynamical equation for the \Couleur{\bf T} tensor should apply to any kind of continuum. This \hypertarget{T-eqs-Universal}{expresses the mass-energy equivalence} and the universality of the gravitation force in the framework of \hyperlink{Medium-by-T}{Assumption ({\it b})}. Thus, we assume that the dynamical equation derived for a dust in the presence of a field of external force density \Couleur{${\bf f}$}, Eq. (\ref{Eq T-f}) with (\ref{b^mu}), is true for a general continuum. 

%%%%%%%%%%%%%%%%%%%%%%%%%%%%%%%%%%%%%%%%%%%%%%%%%%%%%%%%%%%%%%%%%%%%%%%%%%%%%%%%
\subsection{Newton's second law for a general continuum}
%%%%%%%%%%%%%%%%%%%%%%%%%%%%%%%%%%%%%%%%%%%%%%%%%%%%%%%%%%%%%%%%%%%%%%%%%%%%%%%%

Instead of fully inducing it from a dust, the dynamical equation for a general continuum may also be obtained by writing Newton's second law for a volume element of the general continuum. The induction from a dust is then used only to define the ``purely material" energy content of a volume element: for a dust, this is unambiguously defined as $\delta E \equiv  \delta m_0 \gamma_{v} c^{2} = \rho_{0} \gamma_{v}\, c^{2}\delta V $, and one finds easily that this is equal to $T^0_{\ \, 0} \delta V$. The latter expression is thus defined to be the relevant energy to write Newton's second law for a volume element of a general continuum. When writing this, the volume density ${\bf f}'$ of the non-gravitational force must now include the volume density of the {\it internal} forces in the continuum, added to the volume density ${\bf f}$ of the external forces. (Only external forces act on a dust, by definition.) The equation of motion thus obtained coincides with (\ref{Eq T-f})$_2$ for a dust, i.e., when ${\bf f}'={\bf f}$ \cite{A54}. (This latter fact is proved by using the bimetric nature of the theory.)

%%%%%%%%%%%%%%%%%%%%%%%%%%%%%%%%%%%%%%%%%%%%%%%%%%%%%%%%%%%%%%%%%%%%%%%%%%%%%%%%
\section{Maxwell equations in a gravitational field}
%%%%%%%%%%%%%%%%%%%%%%%%%%%%%%%%%%%%%%%%%%%%%%%%%%%%%%%%%%%%%%%%%%%%%%%%%%%%%%%%
%%%%%%%%%%%%%%%%%%%%%%%%%%%%%%%%%%%%%%%%%%%%%%%%%%%%%%%%%%%%%%%%%%%%%%%%%%%%%%%%
\subsection{First group, Lorentz force}
%%%%%%%%%%%%%%%%%%%%%%%%%%%%%%%%%%%%%%%%%%%%%%%%%%%%%%%%%%%%%%%%%%%%%%%%%%%%%%%%
\bi
\item i) Field tensor and first group of Maxwell equations. As usual, we assume that the field tensor \Couleur{$\Mat{F}$} derives from a 4-potential \Couleur{{\bf A}}:

\be\label{Def F}
\Couleur{F_{\mu \nu } \equiv  A_{\nu ,\mu } - A_{\mu, \nu } = A_{\nu;\mu } - A_{\mu ; \nu }}.
\ee 
This is equivalent to assuming:
\bi 

\item that \Couleur{$\Mat{F}$} is antisymmetric (\Couleur{$F_{\mu \nu } = - F_{\nu \mu \, })$};

\item and that the first group of the Maxwell equations is satisfied:

\be\label{Maxwell 1}
\Couleur{F_{\lambda \mu \, ,\nu } + F_{\mu \nu ,\lambda 
} + F_{\nu \lambda ,\mu } = F_{\lambda \mu \, ;\nu } + F_{\mu \nu ;\lambda 
} + F_{\nu \lambda ;\mu } = 0}.
\ee 

\ei
\vspace{1mm}

\item ii) Lorentz force in a gravitational field. The expression of the Lorentz force should:

\bi

\item be a space vector, invariant by the 
transformations $\Couleur{x'^0 = \varphi (x^0)}, \quad \Couleur{x'^i= \psi^i(x^1,x^2,x^3)}$;

\item reduce to that valid in SR, when the gravitational field vanishes.

\ei
This leads straightforwardly \cite{A54} to
\be\label{Lorentz force ETG-vector}
\Couleur{{\bf F} = q \left ( {\bf E} +  {\bf v}\wedge \,\frac{{\bf B}}{c} \right),\quad \quad ({\bf a}\wedge {\bf b})^i \equiv e  ^i_{\ \, jk}\,a^j\,b^k},
\ee
where $e_{ ijk}$ is the usual antisymmetric spatial tensor, whose indices are raised or lowered using the spatial metric $\Mat{g}$ in the preferred reference fluid $\mathcal{E}$. We have $e_{ ijk}=\sqrt{g}\, \varepsilon_{ ijk} $ in spatial coordinate systems whose natural basis is direct, $\varepsilon _{ ijk}$ being the signature of the permutation $(i\,j\,k)$. In Eq. (\ref{Lorentz force ETG-vector}), ${\bf E}$ and ${\bf B}$ are the electric and magnetic spatial vector fields in the reference fluid $\mathcal{E}$, with components 
\be\label{E and B}
\Couleur{E^i\equiv \frac{F^i_{\ \,0}}{\beta }}, \quad \Couleur{B^k\equiv -\frac{1}{2}e^{ijk} F_{ij}}.
\ee

\ei

%%%%%%%%%%%%%%%%%%%%%%%%%%%%%%%%%%%%%%%%%%%%%%%%%%%%%%%%%%%%%%%%%%%%%%%%%%%%%%%%
\subsection{Second group of Maxwell equations in a gravitational field}
%%%%%%%%%%%%%%%%%%%%%%%%%%%%%%%%%%%%%%%%%%%%%%%%%%%%%%%%%%%%%%%%%%%%%%%%%%%%%%%%

\bi

\item i) Equation for the energy-momentum tensor of the electromagnetic field. Let \Couleur{${\bf T}_\mathrm{field}$} be the energy-momentum tensor of the e.m. field \cite{Fock1964}:
\be\label{T em}
\Couleur{T_\mathrm{field}^{\, \mu \nu } \equiv \left (- F^\mu_{\ \ \lambda } F^{\, 
\nu \lambda } + \frac{1}{4}\gamma^{\, \mu \nu } F_{\lambda 
\rho } F^{\lambda \rho \, } \right)/4\pi }. 
\ee
The \textit{total} energy-momentum is assumed to be \Couleur{${\bf T} = {\bf T}_\mathrm{charged\ medium} + {\bf T}_\mathrm{field}$}. Since \hyperlink{T-eqs-Universal}{we assume that Eq. (\ref{Eq T-f}) is universal}, \Couleur{${\bf T}$} obeys Eq. (\ref{Eq T-f}) for continuum dynamics, without any non-gravitational external force. The charged medium obeys Eq. (\ref{Eq T-f}), with the non-gravitational force being the Lorentz force (\ref{Lorentz force ETG-vector}) --- more precisely, with the Lorentz force volume density ${\bf f}\equiv \delta {\bf F}/\delta V$, got by substituting $\delta q = \rho _{\mathrm{el}} \delta V$ for $q$ into (\ref{Lorentz force ETG-vector}), where $\rho _{\mathrm{el}}$ is the volume density of electric charge in the preferred reference fluid. Combining the two using their linearity, we get:
\be\label{Eq T-field}
\Couleur{T_{\mathrm{field}\ \,;\nu}^{0 \nu}=b^0({\bf T}_\mathrm{field})-\frac{{\bf f.v}}{c\beta},
\qquad T_{\mathrm{field}\ \,;\nu}^{i \nu} =b^i({\bf T}_\mathrm{field})- f^i}.
\ee
That is, {\bf the e.m. field behaves as a material continuum subjected to the gravitation and to the opposite of the Lorentz force}.\\

\item ii) Gravitationally-modified second group. Equation (\ref{Eq T-field}), after a short algebra, can be shown \cite{A54} to be equivalent to the following {\it gravitationally-modified second group:}

\be\label{Eq T-field-3}
\Couleur{F^\mu_{\ \ \lambda }\,F^{\lambda \nu }_{\ \ \,;\nu }= 4 \pi b^\mu \left ({\bf T}_\mathrm{field} \right)-4 \pi F^\mu_{\ \ \lambda }\,\frac{J^\lambda }{c}}.
\ee

\vspace{1mm}
Our main comments about this new set of equations are as follows:

\bi

\item If the gravitational field is time-independent in the preferred reference fluid, we have $b^\mu=0$ from (\ref{b^mu}). If moreover the matrix \Couleur{$(F^\mu_{\ \ \nu})$} is invertible, then (\ref{Eq T-field-3}) hence reduces to the gravitationally-modified Maxwell equations in metric theories of gravity:
\be\label{Maxwell GR}
F^{\mu \nu } _{\ \ \,;\nu  }= -4 \pi \frac{J^\mu  }{c}.
\ee
\vspace{1mm}

\item When is the matrix \Couleur{$(F^\mu_{\ \ \nu})$} invertible?  It is easy to check that \Couleur{$\ \ \mathrm{det}\,(F^\mu_{\ \,\nu}) \ne 0$} is equivalent to \Couleur{${\bf E.B}\equiv \Mat{g}({\bf E},{\bf B})\ne 0$}, i.e., one invariant of the field does not vanish. This is true at a generic point for a generic e.m. field, but it is not true for the simplest examples of such fields: purely electric fields, purely magnetic fields, and ``simple e.m. waves".\\

\item In a variable gravitational field such that \Couleur{$b^\mu \ne 0$}, and if the matrix \Couleur{$(F^\mu_{\ \ \nu})$} is invertible, Eq. (\ref{Eq T-field-3}) may lead to {\bf macroscopic charge production/destruction}:
\be\label{Charge rate ETG}
\Couleur{\hat{\rho  } \equiv \left ( J^\mu \right )_{; \mu}= c \left( (\Mat{F}^{-1})^\mu  _{\ \,\nu }\,b^\nu ({\bf T}_\mathrm{field}) \right )_{; \mu}}.
\ee
It had previously been found \cite{A20,A35} that, in a variable gravitational field, this theory predicts macroscopic matter production/destruction in general (i.e. not especially for a charged medium: e.g. for a perfect fluid). It had also been easily shown that the amounts are extremely tenuous in usual situations, hence are compatible with the experimental evidence on mass conservation.  In contrast with this, it is difficult to assess the amounts of charge production/destruction that are predicted in realistic situations. We hope to be able to give estimates soon. By integrating (\ref{Charge rate ETG}) in a spatial domain one finds that the time variation of the charge in that domain happens to be expressible as a surface integral on the boundary of that domain, and that, if the charged system is isolated, then the total charge is conserved. Note that this macroscopic theory does not say which elementary processes would act at the particles' scale. If this phenomenon turns out to be real, it should have some implications for the magnetic fields of the astronomical objects, which currently are not fully understood.

\ei

\ei
%%%%%%%%%%%%%%%%%%%%%%%%%%%%%%%%%%%%%%%%%%%%%%%%%%%%%%%%%%%%%%%%%%%%%%%%%%%%%%%%
\subsection{Consistency with trajectories of photons}\label{photons}
%%%%%%%%%%%%%%%%%%%%%%%%%%%%%%%%%%%%%%%%%%%%%%%%%%%%%%%%%%%%%%%%%%%%%%%%%%%%%%%%

The photon trajectories are defined by Newton's second law (\ref{Newton 2nd law}), with \Couleur{$E=h\nu\quad$} (\Couleur{$\nu\equiv \dd n/ \dd t_{\bf x}$ is the frequency of the e.m. wave as measured with the local time of an observer of the reference fluid, $n$ being the number of periods}), with zero non-gravitational force \Couleur{${\bf F}$}. Therefore, to make the link with the equations for the e.m. field, we have to consider a ``dust of photons". I.e., the energy-momentum tensor of the e.m. field should have the following form:
\be\label{T = tensor product}
\Couleur{T_\mathrm{field}^{\mu \nu } = V^\mu  V^\nu }. 
\ee
This happens to mean exactly that \Couleur{$\Mat{F}$} is a ``null'' field, i.e., the two classical invariants of \Couleur{$\Mat{F}$} are zero. For such a field, one shows (using the bimetric nature of the theory) that Maxwell's second group (\ref{Eq T-field-3}) says exactly \cite{A54}:

\bi

\item that the trajectories of the e.m. energy flux, with velocity \Couleur{${\bf u}$} defined from \Couleur{${\bf T}_\mathrm{field}$} by \Couleur{$T^{0 0}\,u^i=cT^{i 0}$}, are photon trajectories;

\item that one has the continuous form for dust of the energy equation.

\ei

%%%%%%%%%%%%%%%%%%%%%%%%%%%%%%%%%%%%%%%%%%%%%%%%%%%%%%%%%%%%%%%%%%%%%%%%%%%%%%%%
\section{Conclusion}
%%%%%%%%%%%%%%%%%%%%%%%%%%%%%%%%%%%%%%%%%%%%%%%%%%%%%%%%%%%%%%%%%%%%%%%%%%%%%%%%

In the investigated theory of gravitation, particle dynamics is defined by an extension to curved spacetime of the relativistic form of Newton's second law, with the gravity acceleration \Couleur{${\bf g}$} being derived from a spatial potential: Eq. (\ref{Newton 2nd law}) with (\ref{Def vector g}). The equation for continuum dynamics is got by applying (\ref{Newton 2nd law}) to a dust and by postulating that the equation thus obtained, Eq. (\ref{Eq T-f}) with (\ref{b^mu}), is valid for a general continuum or system of fields, i.e., for a general form of the energy-momentum tensor ${\bf T}$.

The Maxwell equations in a gravitational field are obtained by applying the equation for continuum dynamics to the energy-momentum tensor of the e.m. field. They are consistent with photon dynamics as defined by Newton's second law. The theory predicts macroscopic charge production/destruction in a variable gravitational field. This is a dangerous point for the theory, but there may be a relation with the magnetic fields of the astronomical objects.\\

%%%%%%%%%%%%%%%%%%%%%%%%%%%%%%%%%%%%%%%%%%%%%%%%%%%%%%%%%%%%%%%%%%%%%%%%%%%%%%%%
\vspace{2mm}
%\newpage

\end{document}